\documentclass[a4paper]{jpconf}

\usepackage{graphicx}

\usepackage{multirow}
\usepackage{marvosym}
\usepackage{amssymb}
\usepackage{float,caption}
\usepackage{tabularx}
\usepackage[export]{adjustbox}
\usepackage{multicol}
\usepackage{wrapfig}

\begin{document}
\title{Diamond-Like Carbon for the Fast Timing MPGD}


\author{A. Colaleo, G. De Robertis, F. Licciulli, M. Maggi, A. Ranieri, F. Simone, A. Valentini, R. Venditti, P. Verwilligen}
\address{Universit\`a di Bari \& INFN sez. Bari, Via E. Orabona 4, 70125 Bari, Italy}

\author{M. Cesaria, L. Calcagnile, A.P. Caricato, M. Di Giulio, A. Lorusso, D. Manno, M. Martino, A. Perrone, G. Quarta, A. Serra} 
\address{Universit\`a del Salento \& INFN sez. Lecce, Via Per Arnesano, 73100 Lecce, Italy}

\author{M. Ressegotti, C. Riccardi, P. Salvini, P. Vitulo}
\address{Universit\`a di Pavia \& INFN sez. Pavia, Via Agostino Bassi 6, 27100 Pavia, Italy}

\author{I.Vai}
\address{Universit\`a di Bergamo \& INFN sez. Pavia, viale Marconi 5, 24044 Dalmine Bergamo, Italy}

\author{C. Roskas, M. Tytgat}
\address{Universiteit Gent, Proeftuinstraat 86, 9000 Gent, Belgium} 

\ead{piet.verwilligen@ba.infn.it}

\begin{abstract}
The present generation of Micro-Pattern Gaseous Detectors (MPGDs) are radiation hard detectors, capable of detecting efficiently particle rates of several MHz/cm$^2$, while exhibiting good spatial resolution ($\leq 50\,\mu$m) and modest time resolution of 5–10\,ns, which satisfies the current generation of experiments (High Luminosity LHC upgrades of CMS and ATLAS) but it is not sufficient for bunch crossing identification of fast timing systems at FCC-hh. Thanks to the application of thin resistive films such as Diamond-Like Carbon (DLC) a new detector concept was conceived: Fast Timing MPGD (FTM). In the FTM the drift volume of the detector has been divided in several layers each with their own amplification structure. The use of resistive electrodes makes the entire structure transparent for electrical signals. After some first initial encouraging results, progress has been slowed down due to problems with the wet-etching of DLC-coated polyimide foils. To solve these problems a more in-depth knowledge of the internal stress of the DLC together with the DLC-polyimide adhesion is required. We will report on the production of DLC films produced in Italy with Ion Beam Sputtering and Pulsed Laser Deposition, where we are searching to improve the adhesion of the thin DLC films, combined with a very high uniformity of the resistivity values. 
\end{abstract}

\section{Introduction}
\begin{figure}[h]
\includegraphics[width=13pc]{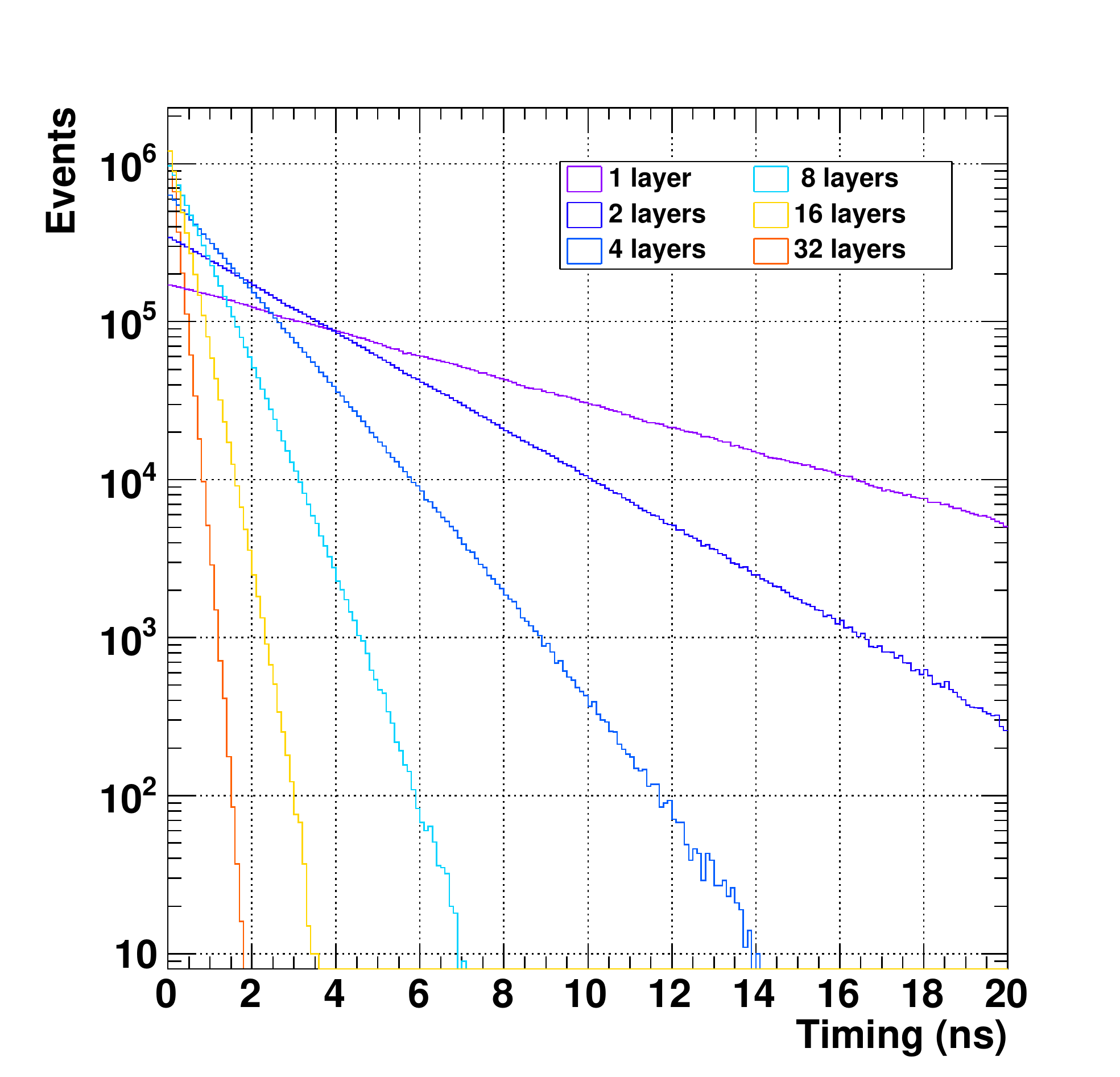}
\includegraphics[width=13pc]{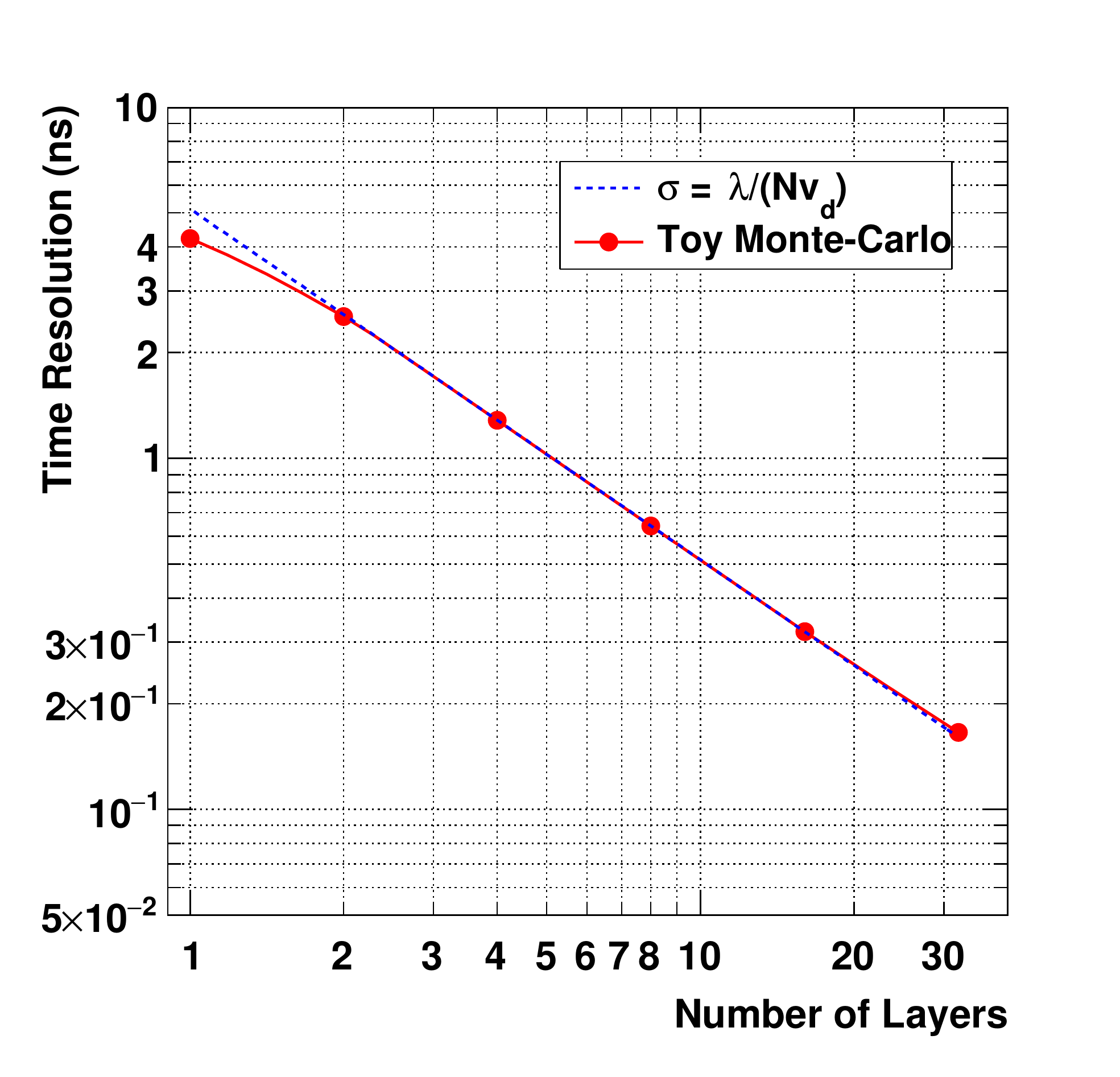}
\includegraphics[width=12pc]{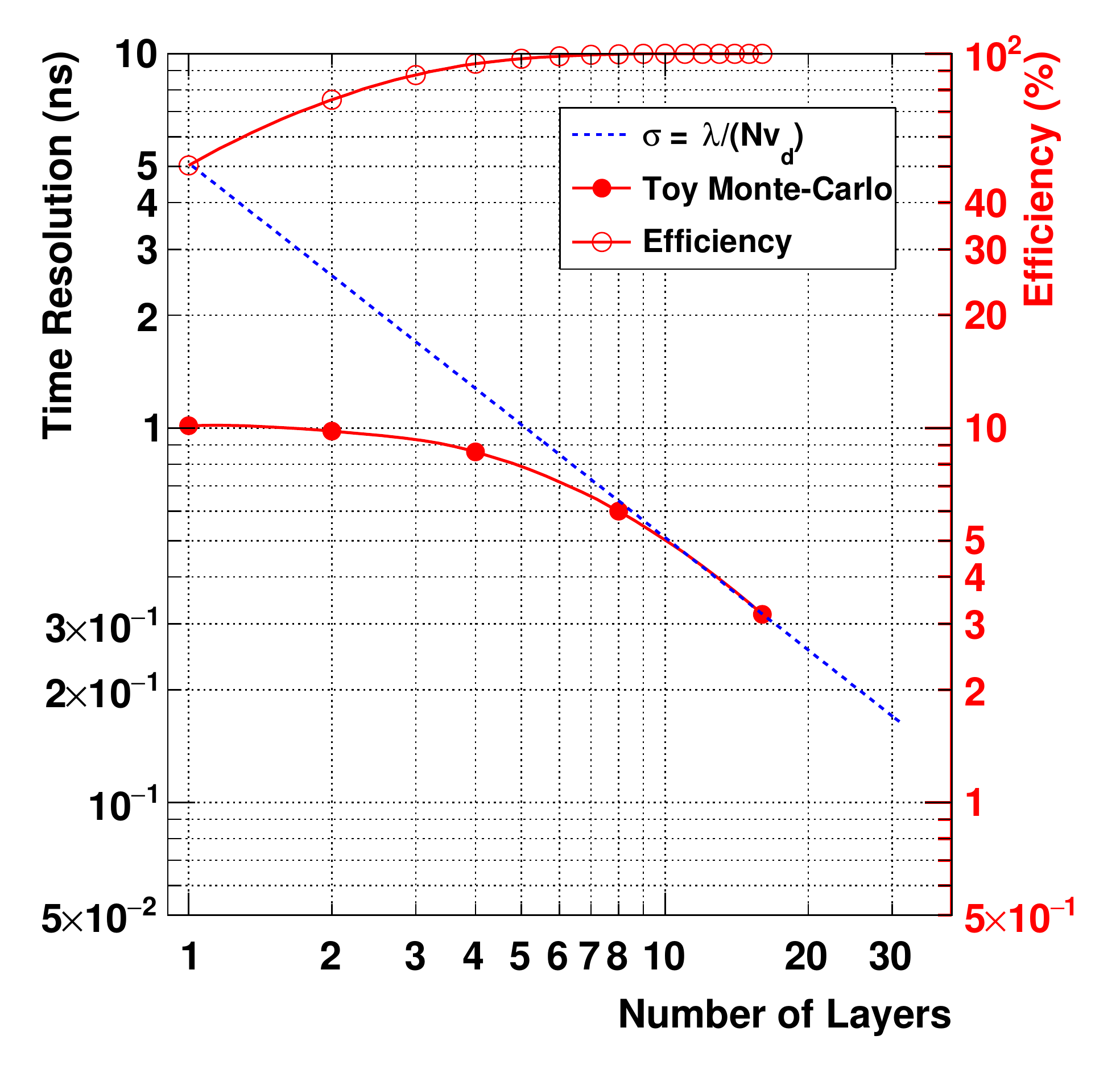}
\caption{\label{lab:ftm-principle}Left: Arrival time (in Ar:CO$_2$ 70:30) for the first electron to the amplification structure for a 4\,mm gas volume split in 1--32 layers. Middle: Time resolution for 4\,mm gas volume split in 1--32 layers. Right: Time resolution and efficiency for 1--16 layers of fixed 250\,$\mu$m width.}
\end{figure}

MPGDs are characterised by micro-metric anode structures that allow for fast collection of ions created during the avalanche process. This leads to a design where there is typically a milimetric gap for creation of primary ion-electron pairs and an amplification structure with the dimension of few tens to hundreds of micrometers. The time resolution of the MPGD is than limited to $\mathcal{O}$(5--10\,ns) due to fluctuations of the primary ionization. Splitting the drift gap then into many smaller gaps, each with its own amplification structure leads to a reduction of the fluctuations in the distance between the closest ion-electron pair and the amplfication structure (see Fig~\ref{lab:ftm-principle} left). Incrementing the number of amplification stages has a negative effect on the cost of such a detector, which can be kept under control by avoiding the single amplfication structures to have also their own readout strips and electronics. If instead one makes a fully resistive detector, signals of any (intermediate) layer can be picked up by external readout strips. This requires fully resistive amplification structures: one possible candidate to replace the metallic GEM electrodes is Diamond-Like Carbon is an amorphous film containing a significant fraction of $sp^3$ hybridization with attractive mechanical and resistive properties~\cite{Grill, Lifshitz}. DLC films can be made with resistivities spanning 10$^2$ to $10^{16}$\,$\Omega$cm.

\subsection{Fast Timing MPGD Principle}
The time resolution of an MPGD with a gaseous drift gap for primary ionisation creation is limited by the fluctuations in the distance between the first cluster and the amplification structure. Its time resolution can be modeled as $\sigma_t \propto \lambda / v_d$ with $\lambda$ the mean distance between two ionisation encounters and $v_d$ the drift velocity of electrons in the gas. It can be shown\footnote{Using the property that the minimum of a set of exponentially distributed observables is also exponentially distributed.} that splitting the large drift gap in several ($N$) micro-gaps, the time resolution improves with a factor $N$: $\sigma_t \propto \lambda / (N v_d)$. This relation is shown with a blue dotted line in Fig~\ref{lab:ftm-principle} middle \& right), while in a red full line a simple monte-carlo experiment confirms this relationship for a 4\,mm gas volume (Ar:CO$_2$ 70:30, $\lambda = 2.8$\,mm$^{-1}$ for a Minimum Ionising Particle (MIP): $\beta \gamma = 3.5$, $v_d = 70$\,$\mu$m/ns at $E=3$\,kV/cm) split in 1 to 32 layers. If on the other hand one considers a detector made of a limited number of fixed 250\,$\mu$m layers, then the time resolution of a single layer starts at $\sim$1\,ns because there is an effective cut-off of signals with long response time due to the maximum drift length of 250\,$\mu$m = 3.6\,ns ($v_d = 70$\,$\mu$m/ns). This is visible in Fig~\ref{lab:ftm-principle} (right), where only at $N = 8$ layers, the time resolution starts following the scaling law discussed above. This improved time resolution (1\,ns w.r.t. 5\,ns expected for $N = 1$) goes obviously at the cost of the detection efficiency that reaches only 100\,\% for $\geq 6$ layers (1.5\,mm total gas volume). 

\subsection{Fast Timing MPGD}
\begin{figure}[h]
\begin{minipage}{14pc}
\includegraphics[width=14pc]{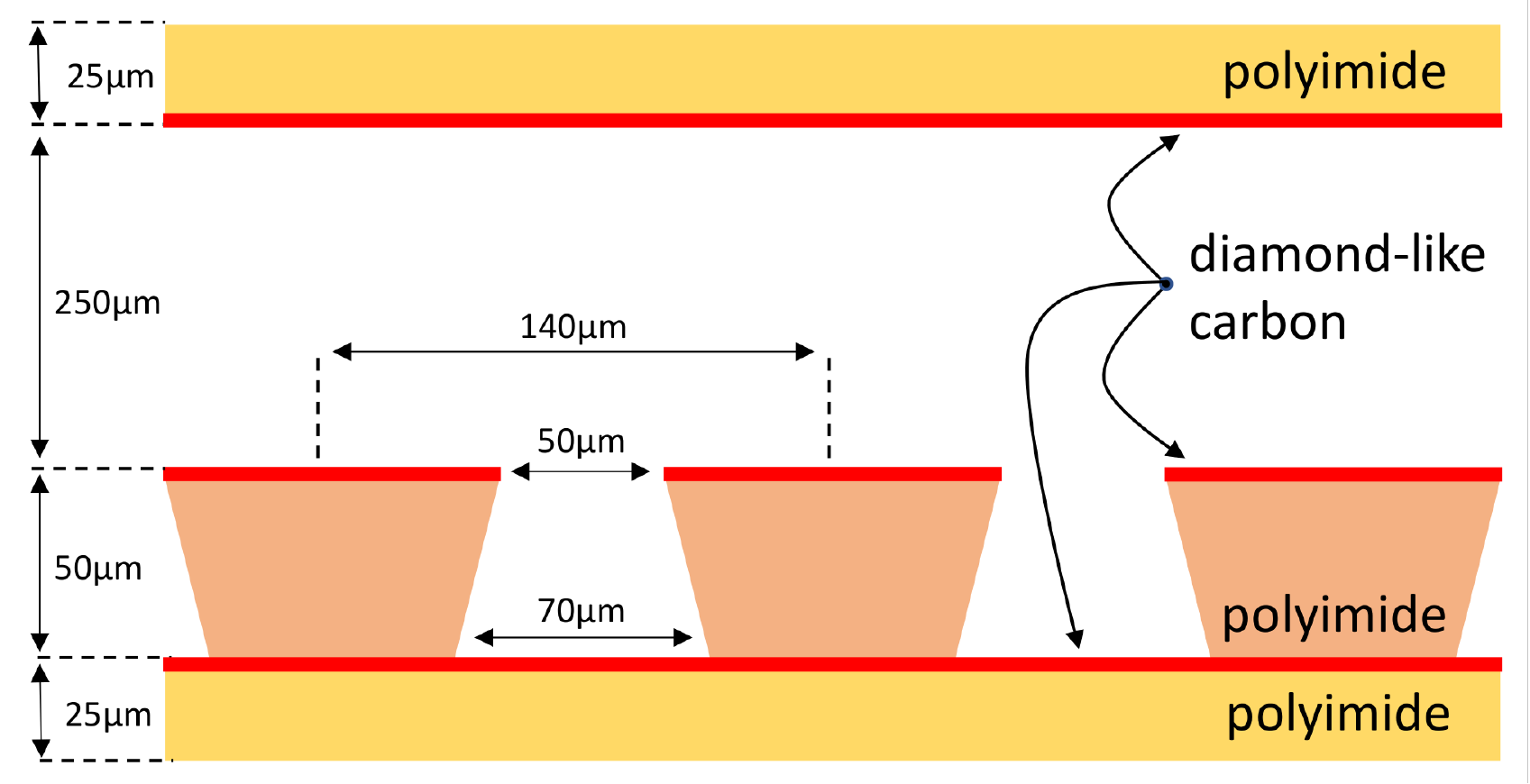}
\caption{\label{lab:ftm-scheme}Schematic view of a single FTM layer.}
\end{minipage}\hspace{2pc}%
\begin{minipage}{14pc}
\includegraphics[width=14pc]{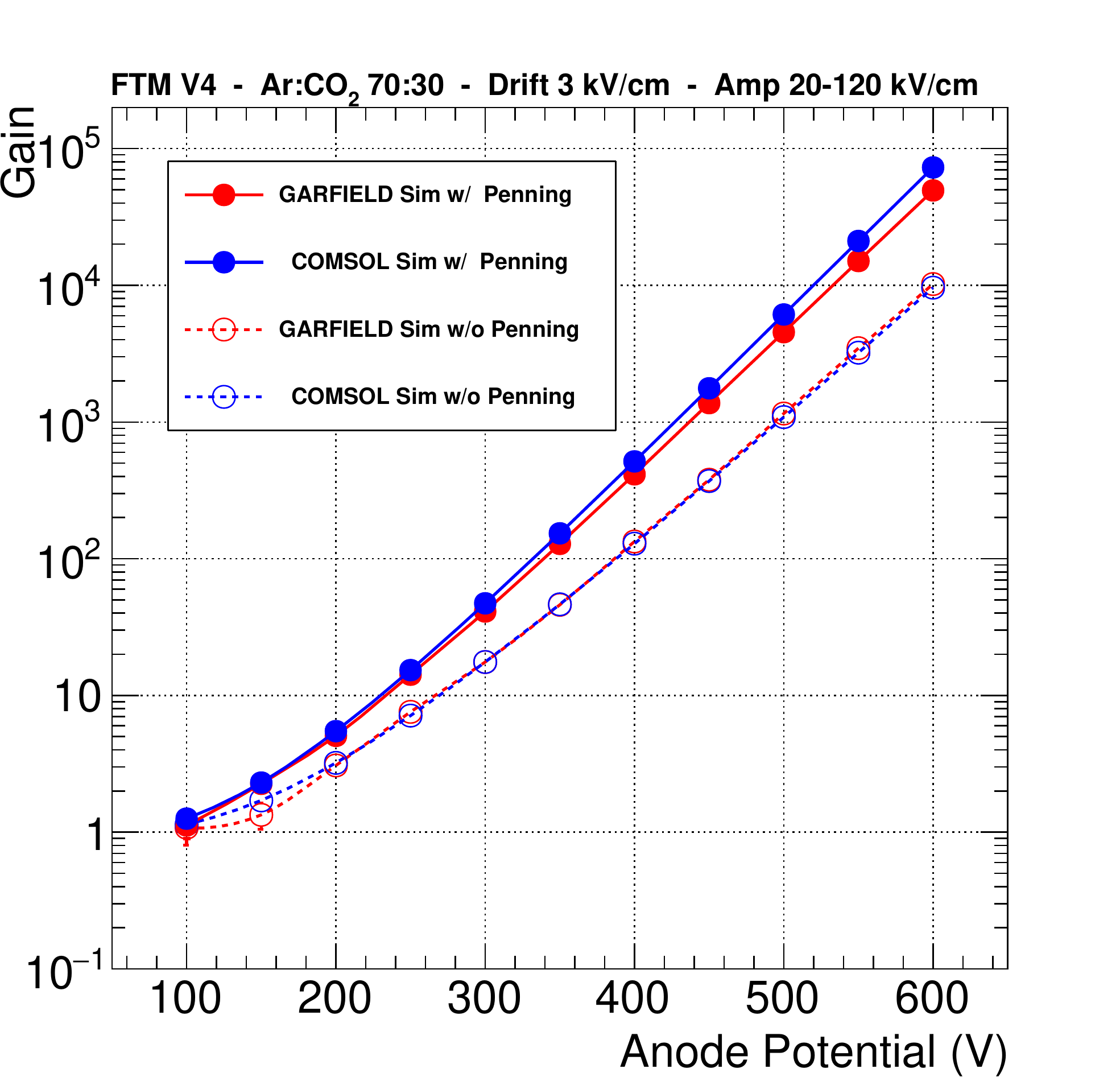}
\caption{\label{lab:ftm-gain}Attained gain in a single FTM layer in Ar:CO$_2$.}
\end{minipage} 
\end{figure}

The Fast Timing MPGD (FTM) was conceived \cite{Maggi} as a $\mu$-well like detector (See Fig~\ref{lab:ftm-scheme}), using 50\,$\mu$m polyimide foils with a GEM-like hole pattern created through the single-mask wet etching process (hexagonal pattern with pitch 140\,$\mu$m, conical holes with 70\,$\mu$m top diameter, 50\,$\mu$m bottom diameter) \cite{Duarte-Pinto}, where the top of the polyimide foil is covered with a $\sim 100$\,nm DLC layer acting as electrode. Simulations of the attained detector gain performed with microscopic tracking (ANSYS \& Garfield$++$) are confirmed by a a simple model that integrates the penning-corrected Townsend ionization coefficient ($\alpha$) \cite{Sahin} along the center of the hole (see Fig~\ref{lab:ftm-gain}). At an electrode potential of 500\,V we expect a gain of $G=10^4$. Thanks to the all resistive nature of the detector sparks are naturally reduced due to the voltage-drop in the resistive electrodes. The choice of the resistivity of the DLC is a trade-off between high signal transparency (high resistivity) and high rate capability (low resistivity). It was choosen to first establish the working principle of the FTM using a rather high resistivity of $\sim100$\,M$\Omega/\Box$, which could then later on tuned to obtain the rate capability necessary for the chosen application.

\subsection{Problems with the current state-of-the-art DLC}
\begin{figure}[h]
\includegraphics[width=14pc]{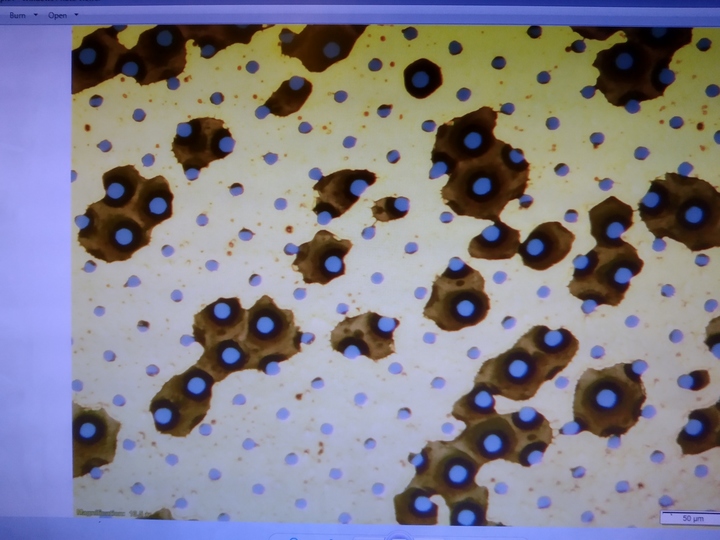}\hspace{2pc}%
\begin{minipage}[b]{14pc}\caption{\label{lab:dlc-delam} Microscopic view of the DLC coated PI foils after the etching process. The DLC (yellow) is locally delaminated, showing the PI (brown). Larger hole diameter is visible in the delaminated area.}
\end{minipage}
\end{figure}

The production of the FTM requires a good knowledge of the chemical wet etching process of DLC coated polyimide foils and high quality DLC film with good adhesion to the polyimide base material. For the production of $\mu$-RWELL the DLC side of the PI is effectively protected because glued on the readout PCB, high quality etching results have been established early on, resulting in detectors that can reach a gain of $10^4$ \cite{Gianni}. For the FTM the DLC side of the PI needs to face the drift gas volume and cannot be glued. Several trials have been made using magnetron sputtered DLC in Japan \cite{Ochi} and China \cite{Zhou} but no satisfactory resuls have been obtained. Fig~\ref{lab:dlc-delam} shows a microscopic zoom of the DLC coated PI foils after the etching process. The DLC is locally delaminated (brown areas), resulting also in holes with much larger diameter, leading to reduced electric fields inside the holes and as a consequence low avalanche gain. 

Little progress was made in the past years to improve the DLC adhesion to the polyimide through magnetron sputtering, while the wet etching process was tuned to obtain the best possible results. The only way forward is to understand and improve the quality of the DLC. More insight and control on the DLC films can be obtained using different deposition techniques: Ion beam deposition and Pulsed Laser deposition, which are just a few among many different Physical Vapor Deposition (PVD) techniques. The main difference that affects the quality of the foil is the energy of the species (ions, atoms) used in the deposition process, which ranges from $\mathcal{O}$(0.1\,eV) for evaporation, over $\mathcal{O}$(0.1--10\,eV) for magnetron sputtering, to $\mathcal{O}$(10--100\,eV) for ion-beam and pulsed laser deposition. The main advantages of the magnetron sputtering are fast deposition on large areas and widespread adoption by industry. Ion beam deposition has the advantage of making good quality film and good adhesion while it can deposit many materials on many (also non conducting) substrates. Pulsed Laser deposition has as main advantage the precise control of the DLC $sp^3/sp^2$-ratio, controlling many independent deposition parameters.

\section{Ion Beam Deposition of DLC}
\subsection{Methods}
At the University of Bari a home-made dual ion-beam system with Kaufman ion sources was used to deposit DLC films on polyimide. The setup, shown in Fig~\ref{lab:ibs-setup} and described in more detail in \cite{Valentini}, contists of two ion sources for sputtering and an assistance ion source. For the deposition of DLC films, only one (main) ion source is operated to bombard a 10\,cm diameter pyrolitic graphite target at an angle of $45^\circ$. A carbon film is deposited on a $6\times6$\,cm$^2$ polyimide substrate. The assistance Ar ion source is focussed directly on the polyimide substrate. Organic compounds are removed from the substrate before the DLC deposition starts. During the deposition the assistance beam effectively compactifies the Carbon ions deposited by the main ion beams hence improving the uniformity and quality of the DLC film. A quartz micro-balance is used to determine the thickness of the film.
\begin{center}
\begin{tabularx}{\textwidth}{*{2}{>{\centering\arraybackslash}X}}
   \centering
\includegraphics[width=16pc]{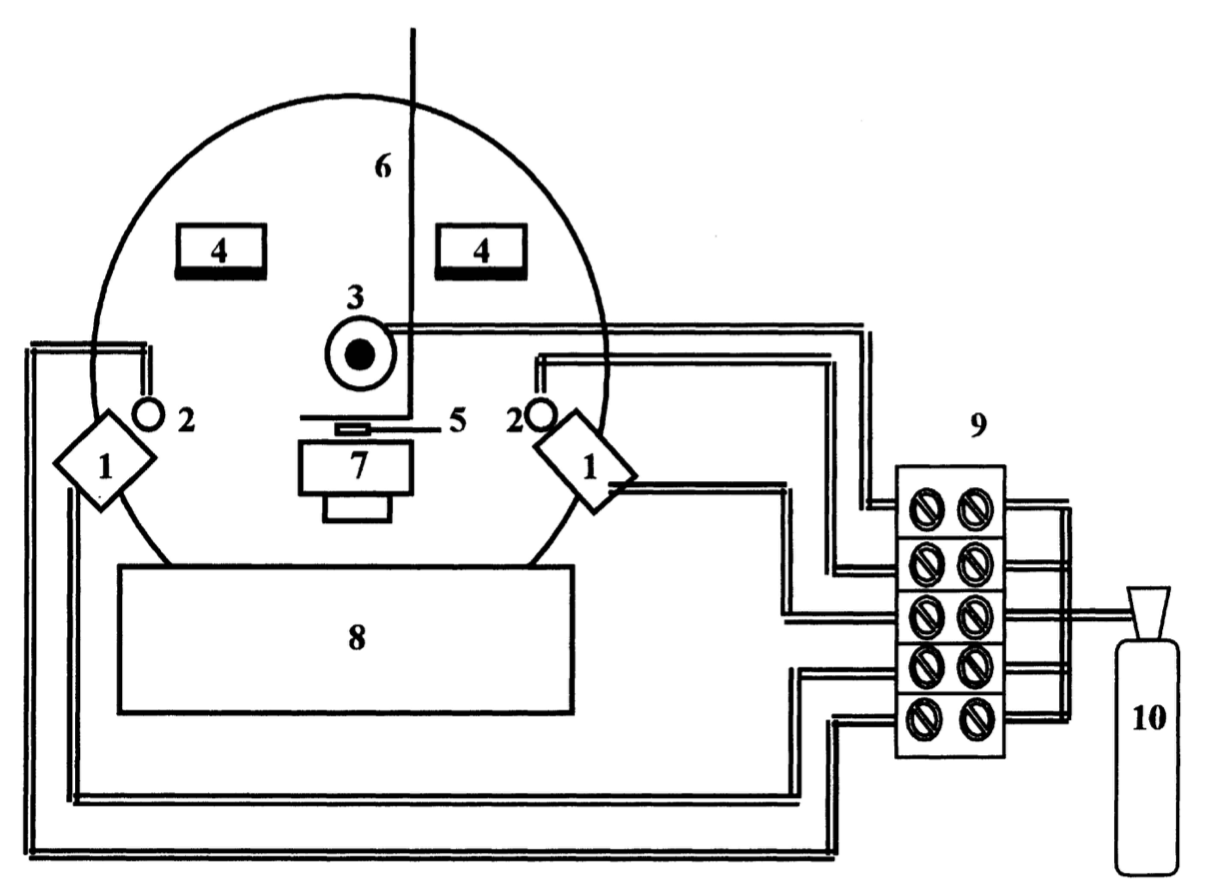}
   \captionof{figure}{\label{lab:ibs-setup} Schematic representation of the dual ion-beam sputtering deposition system used: (1) sputtering ion-beam sources, (3) assistance ion-beam source, (4) targets, (7) substrate holder, (8) turbomolecular pump. Figure taken from \cite{Valentini}. }
& 
\setlength\tabcolsep{4pt}
\vspace{-4cm}
\begin{tabular}{rcccc}
\br
        & 1819 & 1820 & \multicolumn{2}{c}{1821} \\
\hline
\multirow{4}{*}{\rotatebox[origin=t]{90}{\textbf{Main}}} &$Ar$ & $Ar$ & \multicolumn{2}{c}{$Xe$} \\
              & 2.5\,cc/min   & 2.5\,cc/min & \multicolumn{2}{c}{2.5\,cc/min} \\
              & 1200\,eV  & 1200\,eV & \multicolumn{2}{c}{1200\,eV} \\
              & 80\,mA    & 60\,mA   & \multicolumn{2}{c}{60\,mA} \\
\hline
              &$Ar$ & $Ar$ & $Ar$ & $N\bullet$ \\
              & 7\,cc/min   & 7\,cc/min & 5\,cc/min & 2\,cc/min\\
              & 60\,eV   & 100\,eV  & \multicolumn{2}{c}{50\,eV} \\
              & 0.5\,A   & 1.17\,A  & \multicolumn{2}{c}{0.22\,A} \\
\multirow{-5}{*}{\rotatebox[origin=t]{90}{\textbf{Assistance}}} & all time & first 5\,nm & \multicolumn{2}{c}{all time} \\
\br
\end{tabular}
\captionof{table}{Main and Assistance ion-beam settings for the first three DLC samples.\label{lab:table-ibs}}
\end{tabularx}
\end{center}

\subsection{Results}
\begin{figure}[h]
\begin{minipage}{14pc}
\includegraphics[width=16pc]{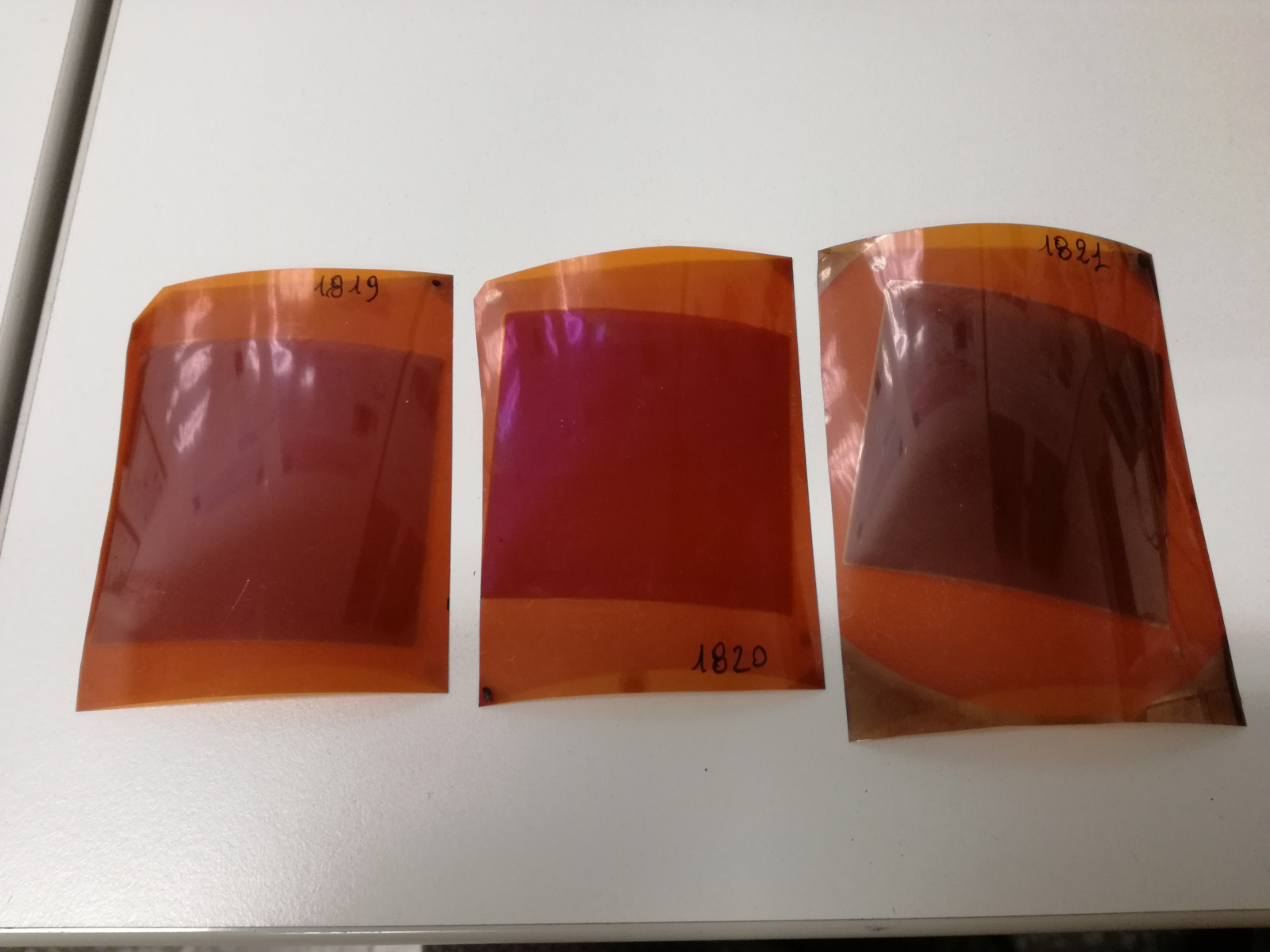}
\caption{\label{lab:ibs-samples} The first three DLC films deposited on polyimide with ion-beam deposition.}
\end{minipage}\hspace{2pc}
\begin{minipage}{14pc}
\includegraphics[width=16pc]{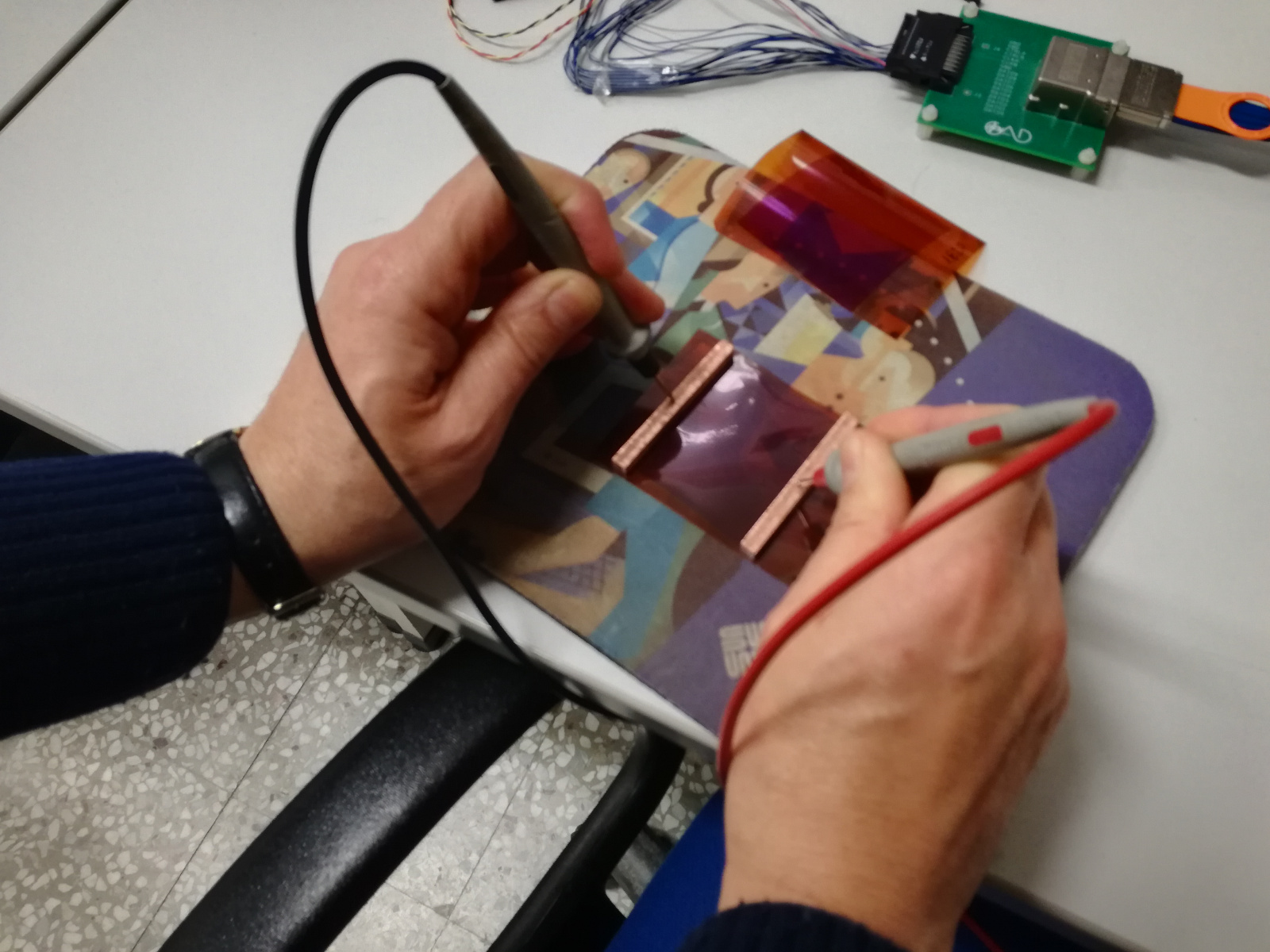}
\caption{\label{lab:ibs-res-meas} Measurement of the surface resistivity with thin Cu-bars and insulation tester.}
\end{minipage}
\end{figure}

Fig~\ref{lab:ibs-samples} shows the first $6\times6$\,cm$^2$ samples were prepared at the end of 2018, varying the settings of the main and assistance ion beams, as detailed in Table~\ref{lab:table-ibs}. The surface resistivity was measured using 0.5\,mm thick Cu bars with 6\,cm length spaced 5\,cm apart, applying a voltage of 50--500\,V with an isolation tester (Megger MIT 410) and measuring the current (Fig~\ref{lab:ibs-res-meas}). To improve the adhesion of the Cu bars with the sample the measurement was performed on a soft mousepad. The resistivity was measured at various voltages and found to be independent of the testing voltage. To remove possible bias we have constructed a tool with concentric circular electrodes with inner electrode radius 1\,cm and outer electrode radius 2\,cm, the measurements agree within uncertainties..

\begin{figure}[h]
\begin{minipage}{16pc}
\includegraphics[width=16pc]{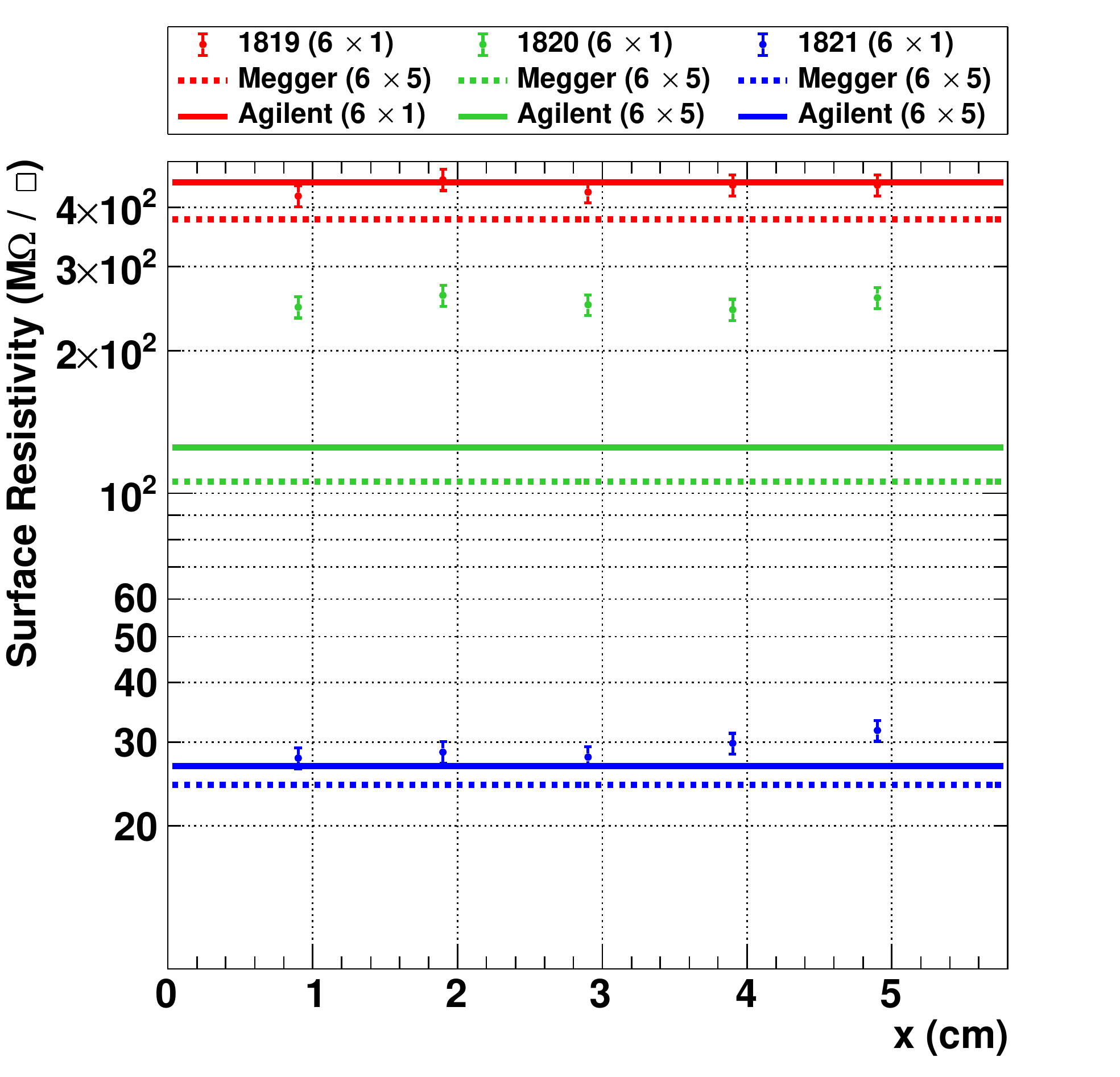}
\caption{\label{lab:ibs-res-unif}Uniformity of surface resistivity of early ion-beam deposition DLC samples.}
\end{minipage}\hspace{2pc}
\begin{minipage}{16pc}
\includegraphics[width=16pc]{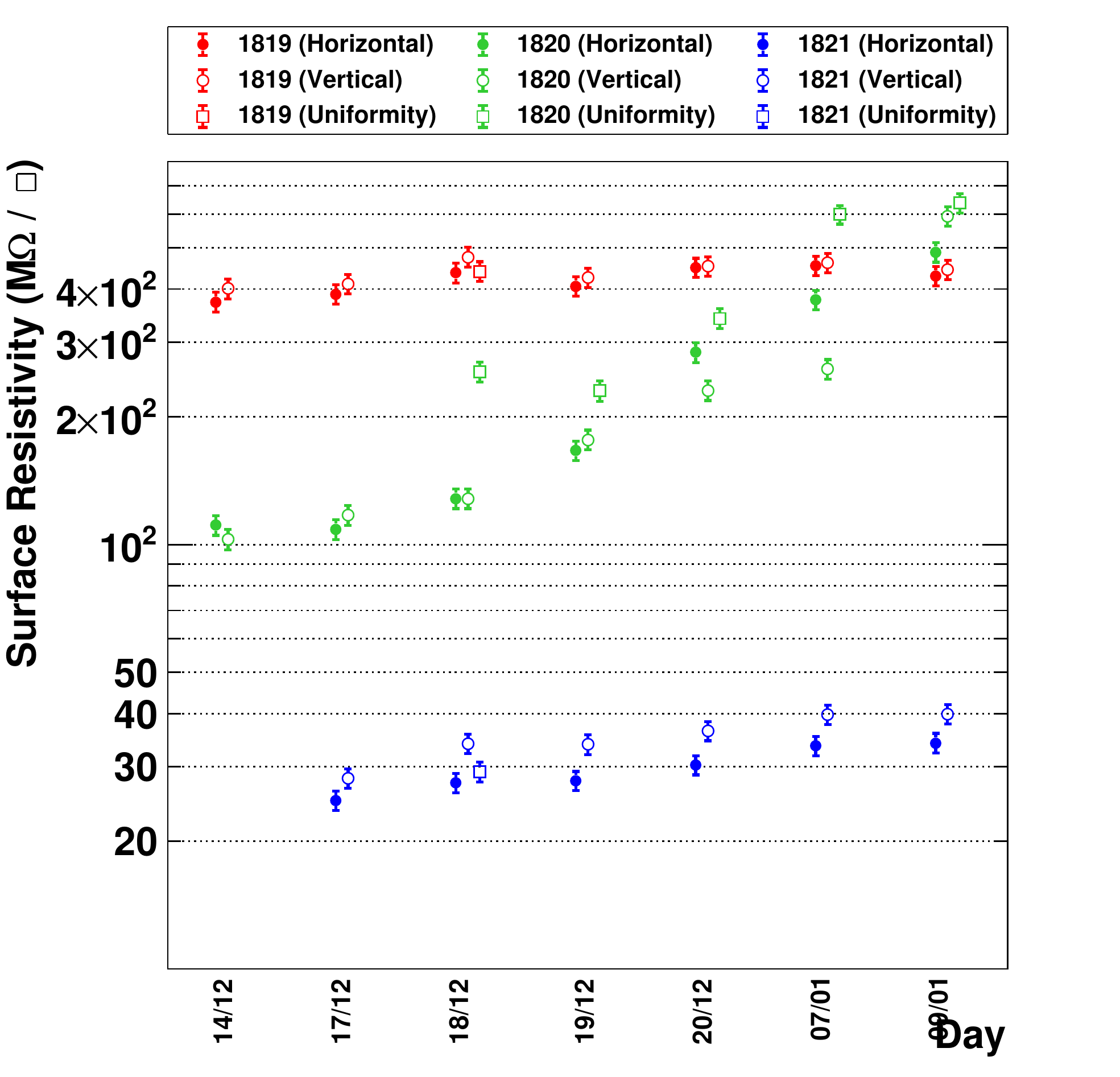}
\caption{\label{lab:ibs-res-time}Time evolution of surface resistivity of early ion-beam deposition DLC samples.}
\end{minipage}
\end{figure}

To asses the uniformity of the surface resistivity the bars were spaced 1cm apart and 5 measurements were taken along the length of the sample. The results are shown in Fig~\ref{lab:ibs-res-unif} where the lines indicate the overall measurement and the markers indicate the uniformity measurements. 

\noindent The first sample (1819) has a resistivity of $\sim 400$\,M$\Omega/\Box$, while the second sample (1820) reached nearly the target of $\sim 100$\,M$\Omega/\Box$, however it's resistivity was not stable in time (uniformity measured after a couple of days). We believe that the limited time of assistance beam lead to a lesser quality film that started oxidating, increasing it's surface resistivity. For the third sample (1821) we introduced Nitrogen doping, leading to a much lower surface resistivity of $\sim 30$\,M$\Omega/\Box$. Further experimentation varying the flux of N$_2$ unfortunately did not allow to tune the surface resistivity and we believe that the inclusion of $N$-ions in the film is already saturated at very low fluxes. This is shown more clearly in Fig~\ref{lab:ibs-res-time} where one can observe the time evolution of the surface resistivity of the first three samples. While the samples deposited with the assistance show a good stability in time, the second sample does not.

\section{Pulsed Laser Deposition of DLC}

\subsection{Methods}
\begin{figure}[h]
\includegraphics[width=14pc]{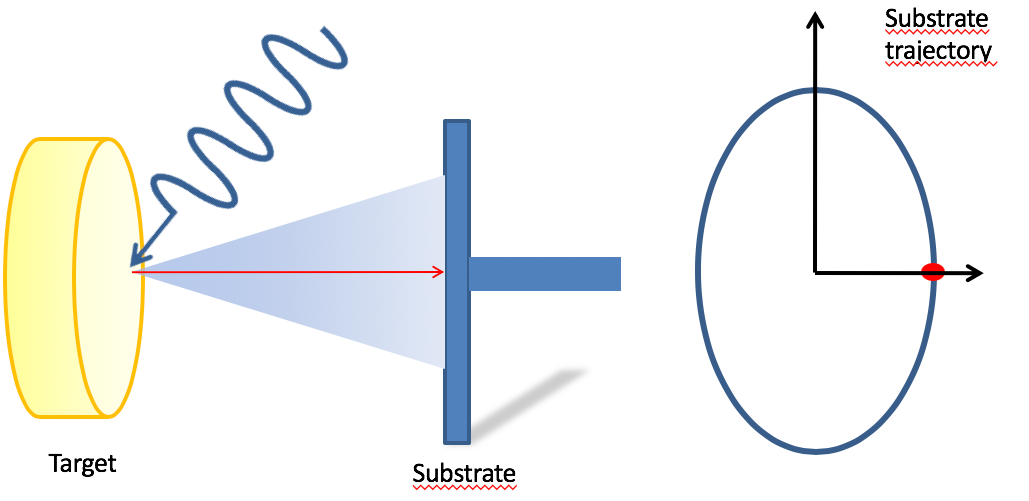}\hspace{2pc}
\begin{minipage}[b]{14pc}\caption{\label{lab:pld-setup} Schematic layout of the Pulsed Laser Deposition with turning substrate holder.}
\end{minipage}
\end{figure}

At the university of Salento, Lecce a pulsed laser deposition system was used to deposit $2\times2$\,cm$^2$ DLC films on polyimide and Si/SiO$_2$ substrates. The laser used for the ablation of graphite is a multi-gas eximer laser (248--193\,nm) with pulses of 20\,ns at 10\,Hz and with pulse energy of max 400\,mJ. The experimental setup is described in full detail in \cite{Caricato}. The laser impings in a vacuum chamber on a rotating pyrolitic graphite target creating a plasma plume containing different chemical species based on carbon atoms (C, C$^+$, C$_2$, C$_2^+$, \ldots). The ablated species impinging on a substrate placed in front of the graphite target that can be fixed or in motion in order to homogenize the deposited film. A sketch of the setup is shown in Fig~\ref{lab:pld-setup}. The resistivity of the DLC films can be tuned varying the laser fluence, as shown in Fig~\ref{lab:pld-fluence}. 

\subsection{Results}
\begin{figure}[h]
\begin{minipage}{16pc}
\includegraphics[width=16pc]{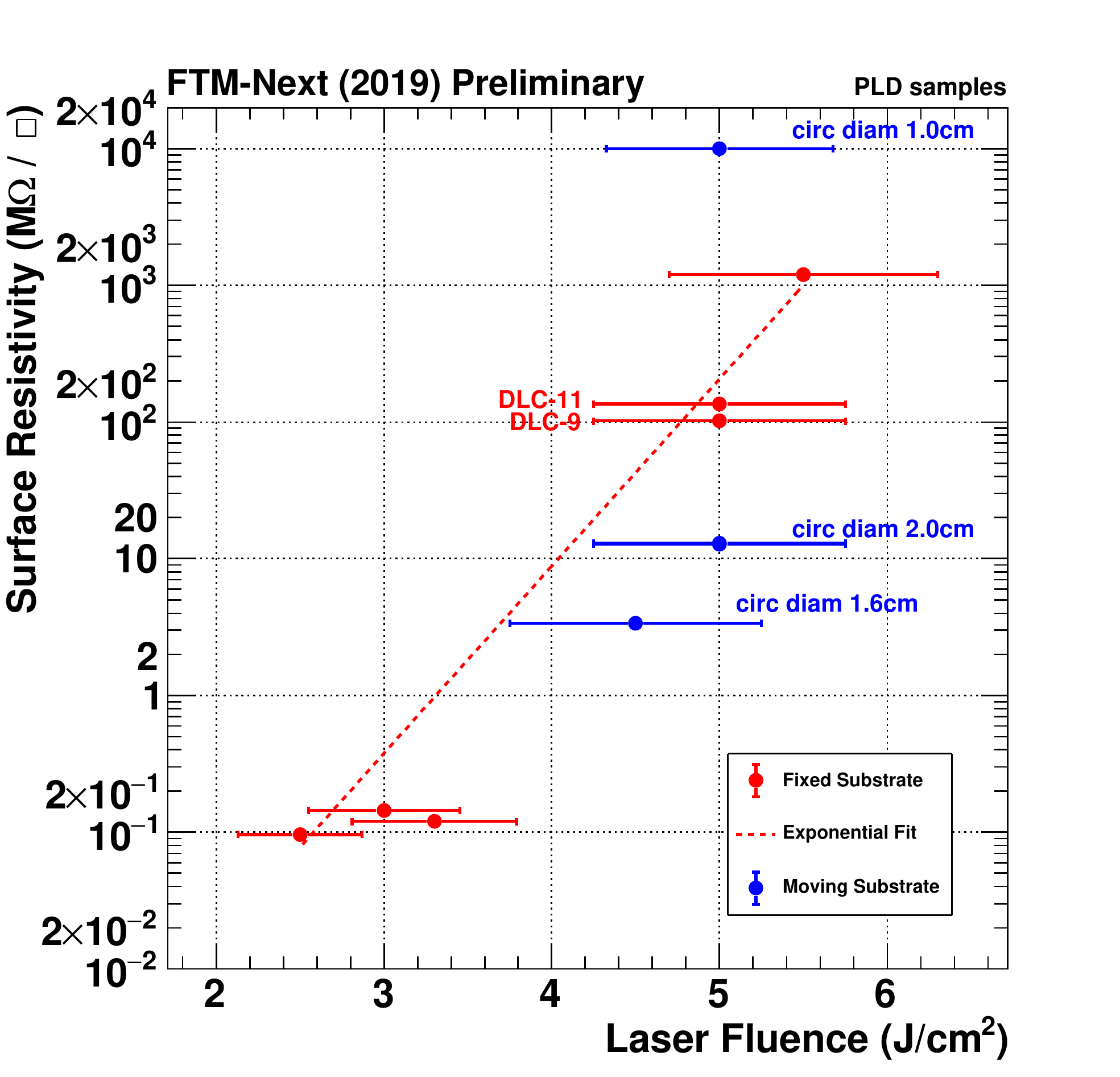}
\caption{\label{lab:pld-fluence} PLD DLC Surface resistivity vs laser fluence for setup without (blue) and with (red) substrate rotation.}
\end{minipage}\hspace{2pc}
\begin{minipage}{18pc}
\includegraphics[width=18pc]{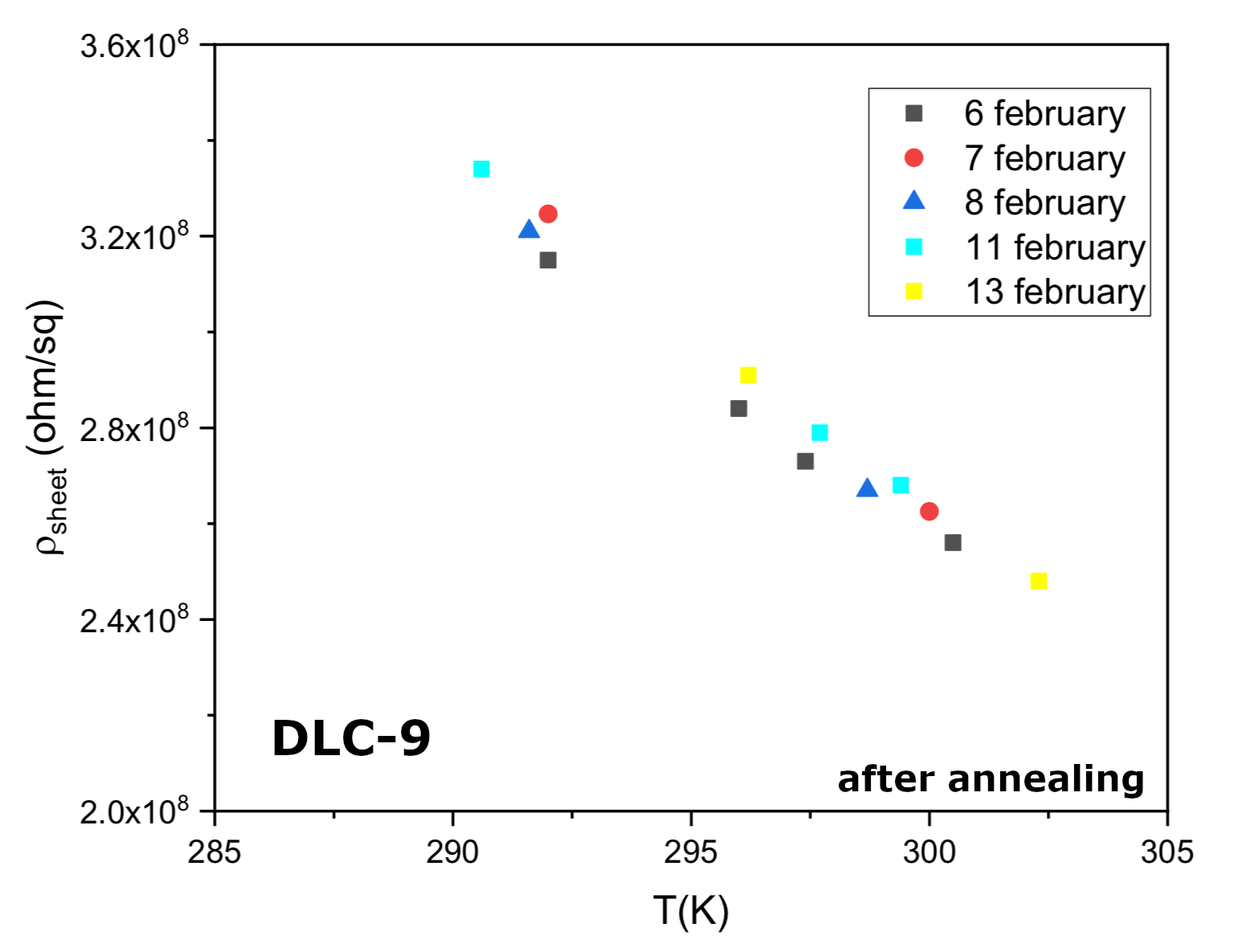}
\caption{\label{lab:pld-temp} PLD DLC Surface resistivity stability and temperature dependence.}
\end{minipage}
\end{figure}

In the deposition by PLD of DLC films, the laser fluence is a critical deposition parameter in order to obtain the desired sheet resistance value. First depositions were performed at different laser fluence values and with different substrates trajectories in order to homogenize the film thickness on an area of $2\times2$\,cm$^2$. The sheet resistance values were found to be dependent on both deposition parameters as can be observed by the red and blue values, respectively, reported in Fig~\ref{lab:pld-fluence} (uncertainty on the laser fuence is $\sim$15\%.). However, reproducible films with an average sheet resistance value of 100\,M$\Omega/\Box$ were obtained at a fuence of 5\,J/cm$^2$  with a fixed substrate. The samples presented an extraordinary stability in time as can be seen from Fig~\ref{lab:pld-temp} where the surface resistivity is reported as function of temperature measured in a time-span of one week. These measurements were performed with a four-point probe stations using the Van Der Pauw method and after annealing at a temperature of 150$^\circ$C,, which changed substantially the surface resistivity from 100\,M$\Omega$/$\Box$ to $\sim 300$\,M$\Omega$/$\Box$. The van der Pauw method allows the measurement of the specific resistivity of arbitrairly shaped flat samples, with only condition that the sample should not have isolated holes (i.e. it should be sigly connected). While with this method sub-\% uncertainties on the resistivity value can be reached, the central value was fluctuating. Therefore the RMS on 50\, individual measurements was taken to estimate the uncertainty on the resistivity, being $\sim10$\%.

\begin{figure}[h]
\includegraphics[width=14pc]{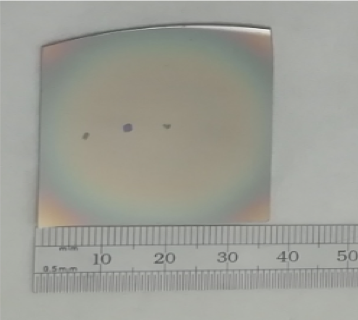}\hspace{2pc}
\begin{minipage}[b]{14pc}\caption{\label{lab:pld-sample} Sample of pulsed laser deposition of DLC ($\varnothing \geq 2$\,cm) on a $3\times3$\,cm$^2$ polyimide substrate.}
\end{minipage}
\end{figure}

However later characterization of the samples revealed a non-uniformity of the films due to peculiar characteristics of the plasma plume which was found to be V-shaped \cite{Ursu}. To mitigate this effect the laser spot size was reduced by a factor of 4 and the substrate was simply rotated. Very uniform films were obtained on an area of $3\times3$\,cm$^2$ (Fig~\ref{lab:pld-sample}). Further tests are needed to determine the exact settings to reach the 100\,M$\Omega$/$\Box$ target using these new depostion conditions. Furthermore Raman spectroscopy, X-ray photoelectron spectroscopy and electrical characterizations are in progress to investigate the $sp^3/sp^2$ fraction of the DLC and carrier transport properties in order to have a better control on film quality. 

\section{Outlook} 
Recently activities have started at INFN Bari and Lecce to produce DLC films with high quality and good adhesion on the polyimide substrate to be used in resistive MPGD detectors. Both with the ion-beam setup as with the pulsed laser deposition, the target surface resistivity of 100\,M$\Omega$/$\Box$ has been reached. Further characterization with Raman and XPS is ongoing to investigate in depth the properties of the films. The next step will consist of testing the wet-etch procedure to perforate the foils with a GEM-like mask. The PLD deposition will enlarge the samples to $2\times2$\,cm$^2$ such that these can be used for the construction of a small-size FTM prototype. Both techniques will try to cover the DLC with a Cu mask such that double-mask GEM etching can be performed.




\ack The author would like to thank INFN CSN-V for giving the possibility to develop the FTM project, and all colleagues for their enduring encouragements.

\end{document}